\documentclass{article}
\usepackage{spconf,amsmath,graphicx,}
\usepackage{amssymb,bm,bigstrut,subfigure,multirow,makecell,cite}
\usepackage{threeparttable}

\title{A QP-adaptive Mechanism for CNN-based Filter in Video Coding}
\name{Chao Liu$^{\ast}$, Heming Sun$^{\dagger \ddagger}$, Jiro Katto$^{\dagger }$, Xiaoyang Zeng$^{\ast}$ and Yibo Fan$^{\ast}$ \thanks{Corresponding author: Heming Sun, hemingsun@aoni.waseda.jp}}

\address{\small{$^{\ast}$State Key Lab of ASIC and System,
Fudan University, Shanghai P.R. China }
\small{$^{\ddagger}$JST, PRESTO, 4-1-8 Honcho, Kawaguchi, Saitama, Japan} \\
 \small{$^{\dagger}$Waseda Research Institute for Science and Engineering, Waseda University, Tokyo Japan}\\
 \small{$^{\star}$Graduate School of Fundamental Science and Engineering, Waseda University, Tokyo Japan}}

 \address{\normalsize{$^{\ast}$Fudan University, Shanghai, P.R. China; }
 \normalsize{$^{\dagger}$Waseda University, Tokyo, Japan; }
 \normalsize{$^{\ddagger}$JST, PRESTO, Saitama, Japan} }

\begin{document}
%
\maketitle
\small
\begin{abstract}
Convolutional neural network (CNN)-based filters have achieved great success in video coding.
However, in most previous works, individual models are needed for each quantization parameter (QP) band.
This paper presents a generic method to help an arbitrary CNN-filter handle different quantization noise.
We model the quantization noise problem and implement a feasible solution on CNN, which introduces the quantization step (Qstep) into the convolution.
When the quantization noise increases, the ability of the CNN-filter to suppress noise improves accordingly.
This method can be used directly to replace the (vanilla) convolution layer in any existing CNN-filters.
By using only 25\% of the parameters,
the proposed method achieves better performance than using multiple models with VTM-6.3 anchor. Besides, an additional BD-rate reduction of 0.2\% is achieved by our proposed method for chroma components.


\end{abstract}
\section{Introduction}
\label{sec:intro}

Quantization \cite{quantization} in the hybrid coding framework like H.265/HEVC \cite{h265} and H.266/VVC \cite{h266} is a crucial part of the lossy compression. However, it also causes some severe distortion and artifacts like ringing and Gibbs effects. Filters such as deblocking (DB), sample adaptive offset (SAO), and adaptive loop filter (ALF) are proposed to alleviate these artifacts. Also, learning-based filters, especially CNN-based filters\cite{liu2019dual, mine, VRCNN, DCAD, Tucodec, sun2020image, klopp2020utilising, li2019loop, he2018enhancing, zhang2018residual, jia2019content, li2017cnn}, have shown great potential and arouse widespread interest. Previous studies show that different structures and designs such as serial network\cite{mine, DCAD, Tucodec, sun2020image, he2018enhancing, li2017cnn}, parallel network\cite{liu2019dual, VRCNN, li2019loop, zhang2018residual, jia2019content} can make significant improvements to both subjective and objective qualities. Liu \textit{et al.} \cite{mine} proposed to use depth separable convolution (DSC\cite{DSC}) in CNN to reduce the complexity. Besides, Dai \textit{et al.} proposed a parallel network structure VRCNN \cite{VRCNN}, which uses different sizes of convolution kernels in the same layer to extract the features from different receptive fields. Conversely, Wang \textit{et al.} proposed a DCAD \cite{DCAD} with a serial structure, which stacks 10 convolutional layers, also achieves good performance. Apart from what is mentioned above, Zhou \textit{et al.} proposed to use the res-net\cite{resnet} as the backbone of the proposed Tucodec \cite{Tucodec}, and used leaky-ReLU \cite{LReLU} instead of the ReLU \cite{ReLU} as the activation function.

Although CNN-based in-loop filters have achieved great success. Few previous studies have explored the generalization capabilities of different quantization parameters (QPs) and most works require training a specific model for each QP band. It is impossible to train quantities of models in actual filtering due to limited storage resources.
Also, those CNN filters do not make full use of the side information of coding, such as the corresponding QP of the reconstructed image, etc.

In this paper, we propose a novel method to solve this important but easily neglected problem.
Specifically, from the frequency domain, we model this problem and obtain a feasible solution of making a simple filtering model adapt to different quantization noises. By further decomposition, the simplified solution is applied to each convolution layer instead of the first one, so it has great robustness and performance.
With VTM-6.3 anchor, we conduct extensive experiments on four models of different complexity to demonstrate the versatility of the proposed method.
Compared with using multiple models, using a single model with the proposed method achieves about three-quarters reduction on the number of parameters and extra 0.2\% BD-rate reductions on chroma components.

\section{Literature Review and Analysis}
To our best knowledge, only one approach of using QP map \cite{song2018practical} as the extra input has been proposed to solve this problem.
And similar methods \cite{zhang2019recursive, zhu2020residual} have been proposed based on the QP map.
The network \cite{song2018practical} can better control the filtering strength by using the internal relationship between the QP map and the distortion. To analysis its working principle, we consider the first convolution layer of this model:
\begin{equation}\label{eq:ref1}
  \hat{\bm{y}} = \bm{w}\ast\{\hm{y};QP\}+\bm{b}
\end{equation}
The notations $\bm{y}$ and $\hat{\bm{y}}$ are the input and the output, $\bm{w}$ and $\bm{b}$ are the weights and the biases, $\ast$ and $\cdot$ are the convolution and the multiplication, $\{\bm{\cdot};\bm{\cdot}\}$ is the concatenation operation. By expanding it:
\begin{align}\label{eq:ref2}
  \hat{\bm{y}} &= \bm{w}_1\ast \hm{y}+\bm{w}_2\ast QP+\bm{b} \notag \\
    &= \bm{w}_1\ast \hm{y}+(\Sigma \bm{w}_2 \cdot QP +\bm{b}) \notag \\
    &= \bm{w}_1\ast \hm{y}+\bm{b'}(QP)
\end{align}
where
\begin{equation}\label{eq:ref3}
  \bm{b'}(QP) = \Sigma \bm{w}_2 \cdot QP +\bm{b}
\end{equation}
From (\ref{eq:ref3}), it can be found that the adaptiveness of \cite{song2018practical} is actually achieved by adding QP to the bias term with a linear function. There are some drawbacks of this method.
1. Using linear models for bias may not explain the internal relationship between QP and filtering strength. 2. This method may be less effective because it is built with the bias. It is the weight rather than bias that dominates CNN, so building the QP-adaptive method for weight may be more effective. 3. It lacks robustness and does not fully tap the potential of the QP, since only the input introduces QP.  Considering these shortcomings, a better adaptive filtering strategy is designed in this paper.

\section{Proposed Method}
In this section, the proposed QP-adaptive mechanism is introduced. To begin with, we present the modeling to the problem. Then the implemented solution for CNN is provided.

\subsection{Proposed QP-adaptive Mechanism}
Given a simple filtering model(we focus on the weight, so the bias is dropped here.):
\begin{equation}\label{eq:1}
  \bm{w} \ast \bm{y} = \hat{\bm{x}}
\end{equation}
where $\bm{w}$ is the trained convolution kernel, $\bm{y}$ is the distorted image, and $\hat{\bm{x}}$ is the filtered image. It is known that the spatial domain convolution is essentially equivalent to frequency domain multiplication.
\begin{equation}\label{eq:2}
  \mathcal{F}(\bm{w}) \mathcal{F}(\bm{y}) = \mathcal{F}(\hat{\bm{x}})
\end{equation}
where the notation $\mathcal{F}$ represents the Fourier transform.
We assume that this simple model can effectively remove the specific noise in $\bm{y}$ but can't handle variable quantization noise. So it is approximated as the original image $\bm{x}$.
\begin{equation}\label{eq:apr}
  \mathcal{F}(\hat{\bm{x}}) \approx \mathcal{F}(\bm{x})
\end{equation}
As is known, the increase of the coding parameter QP represents the added noise that related to Qstep in the frequency domain. Here $\bm{\epsilon}$ represents the noise caused by the change of QP.
\begin{equation}\label{eq:3}
  \bm{w}'\ast(\bm{y}+\bm{\epsilon}) = \hat{\bm{x}}'
\end{equation}
Therefore, a feasible solution to achieve the adaptiveness for various quantization noise is to find a changeable convolution kernel $\bm{w}'$ to minimize the loss function between the filtered image $\hat{\bm{x}}'$ and the original image $\bm{x}$.
Similarly, (\ref{eq:3}) in frequency domain can be expressed as:
\begin{equation}\label{eq:4}
  \mathcal{F}(\bm{w'})(\mathcal{F}(\bm{y})+\mathcal{F}(\bm{\epsilon})) = \mathcal{F}(\hat{\bm{x}}')
\end{equation}
Here we choose classical mean square error (MSE) as the loss function.
\begin{equation}\label{eq:5}
\mathcal{L} =\mathbb{E}|\bm{x} - \hat{\bm{x}}'|^2
\end{equation}
where $\mathbb{E}$ is the notation of expectation. Considering (\ref{eq:apr}) and Parseval's theorem for the Fourier transform, the loss can be transformed into:
\begin{align}\label{eq:6}
    \mathcal{L} =  & \mathbb{E}|\mathcal{F}(\bm{x}) - \mathcal{F}(\hat{\bm{x}'})|^2 \notag \\
       =    & \mathbb{E}|\mathcal{F}(\bm{x}) - \mathcal{F}(\bm{w}') (\mathcal{F}(\bm{y})+\mathcal{F}(\bm{\epsilon}))|^2 \notag \\
    \approx & \mathbb{E}|\mathcal{F}(\bm{x}) - \mathcal{F}(\bm{w}') \left[\mathcal{F}(\bm{x})/\mathcal{F}(\bm{w})+\mathcal{F}(\bm{\epsilon})\right]|^2
\end{align}
By taking the derivative w.r.t. $\mathcal{F}(\bm{w}')$, we can obtain the solution:
\begin{equation}\label{eq:7}
  \mathcal{F}(\bm{w}')= \underbrace{\mathcal{F}(\bm{w})}_{\text{org. filter}} \underbrace{\left[\frac{1} {1+|\mathcal{F}(\bm{w})|^2\mathcal{F}(\bm{n})/\mathcal{F}(\bm{s})}  \right]}_{\text{influence factor}}
\end{equation}
where $\mathcal{F}(\bm{n}) = \mathbb{E}|\mathcal{F}(\bm{\epsilon})|^2 $ and $\mathcal{F}(\bm{s}) = \mathbb{E}|\mathcal{F}(\bm{x})|^2$. Here, the first term $\mathcal{F}(\bm{w})$ is the original filter in (\ref{eq:2}), and the second term is equivalent to the influence factor that compensats for the increased quantization noise.
It can be found that this solution is similar in form to Wiener deconvolution \cite{wienerdeconv}. The difference lies in the motivations and forms. Wiener deconvolution hopes to recover the original signal from the distorted signal by using the priors of the input signal, noise, and degradation function. 
While this solution doesn't have the concept of degradation function and it aims at making a specific filter become adaptive to the changing quantization noise.

\begin{table*}[tbp]
  \centering
  \small
\begin{threeparttable}
  \caption{\small The Comparison of Different Multi-QP strategies on Overall BD-rate Reductions and the Number of Model Parameters} \label{tab:bdrate}%
    \setlength{\tabcolsep}{1.1mm}{
    \begin{tabular}{c|c|r|r|r|c|r|r|r|r|r|r|r}
    \hline
    \multirow{2}[4]{*}{Models} & \multicolumn{4}{c|}{Global (single model $\times$1)} & \multicolumn{4}{c|}{Separate (single model$\times$4)} & \multicolumn{4}{c}{Proposed (single model $\times$1)} \bigstrut\\
\cline{2-13}      & Param. & \multicolumn{1}{c|}{Y} & \multicolumn{1}{c|}{U} & \multicolumn{1}{c|}{V} & Param. & \multicolumn{1}{c|}{Y} & \multicolumn{1}{c|}{U} & \multicolumn{1}{c|}{V} & \multicolumn{1}{c|}{Param.} & \multicolumn{1}{c|}{Y} & \multicolumn{1}{c|}{U} & \multicolumn{1}{c}{V} \bigstrut\\
    \hline
    Liu \textit{et al}\cite{mine} & 12,266 & -0.97\% & -1.55\% & -2.61\% & 12,266$\times$4 & \textbf{-2.28\%} & -1.64\% & -2.68\% & 12,555 & \textbf{-2.28\%} & \textbf{-1.70\%} & \textbf{-2.90\%} \bigstrut\\
    \hline
    VRCNN\cite{VRCNN} & 54,512 & -0.48\% & -1.57\% & -2.23\% & 54,512$\times$4 & \textbf{-1.88\%} & -1.58\% & -2.34\% & 54,673 & -1.85\% & \textbf{-1.62\%} & \textbf{-2.46\%} \bigstrut\\
    \hline
    DCAD\cite{DCAD} & 296,641 & -2.21\% & -2.00\% & -3.07\% & 296,641$\times$4 & \textbf{-3.83\%} & -2.46\% & -3.84\% & 297,218 & -3.74\% & \textbf{-2.78\%} & \textbf{-3.93\%} \bigstrut\\
    \hline
    Tucodec\cite{Tucodec} & 447,681 & -3.72\% & -2.73\% & -3.63\% & 447,681$\times$4 & -4.49\% & -2.72\% & -3.88\% & 448,514 & \textbf{-4.54\%} & \textbf{-2.95\%} & \textbf{-4.21\%} \bigstrut\\
    \hline
    \end{tabular}%
    }
\end{threeparttable}
\end{table*}%

\begin{figure}[tbp]
  \centering
  \includegraphics[scale = 1]{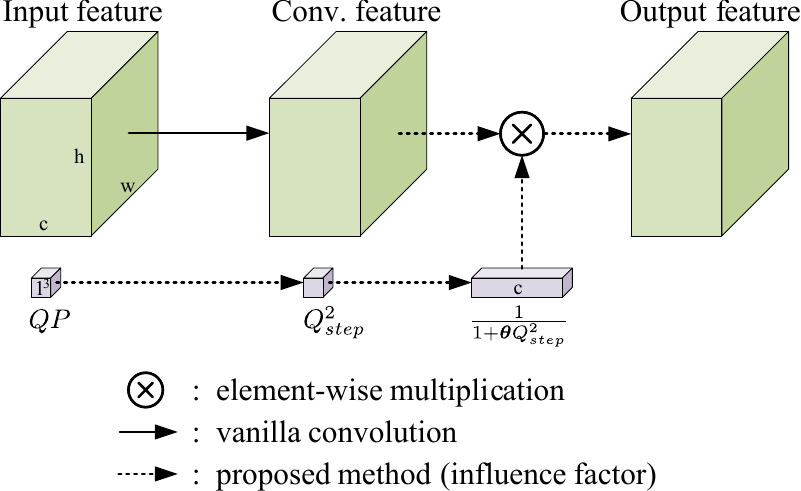}
  \caption{\small Schematic diagram of the proposed method. The influence factors $ 1/(1+\bm{\theta}Q_{step}^2)$are broadcasted to c$\times$ h$\times$ w during the element-wise multiplication, where c, h, w are the channel, height, and weight respectively.}\label{fqpm}
\end{figure}

\begin{figure*}[tbp]
  \small
  \centering
  \includegraphics[scale=0.43]{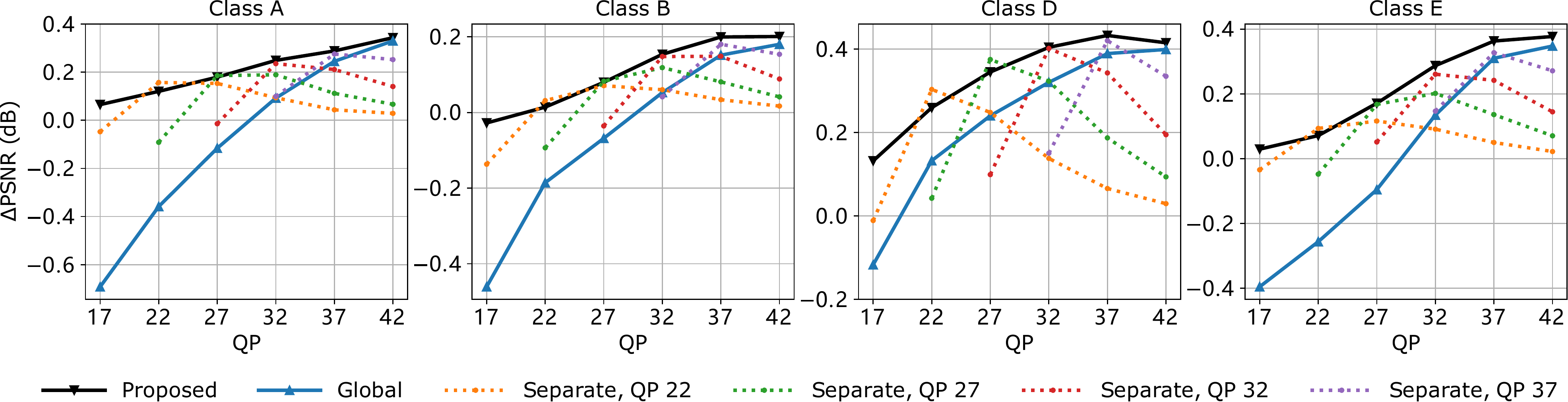}
  \caption{\small Relative PSNR gain curves than VTM-6.3 Anchor. Tucodec is used as the backbone of this test.} \label{fig:psnr}
\end{figure*}

\subsection{Applying QP-adaptive Mechanism to CNN}
From the perspective of the frequency domain, the extracted features from CNNs are equivalent to specific selections of the input image at different frequencies, which establishes the relationship between CNN and the frequency domain in the solution. For example, the Gaussian kernel is a low-pass filter and the Laplacian kernel is a high-pass one. The filter $\bm{w}'$ of the entire frequency band can be decomposed into different sub-filters, and each sub-filter $w'_i$ works in a selected frequency sub-band.
With (\ref{eq:7}), the $\mathcal{F}(\bm{w}')$ could be written as:

\begin{align}\label{eq:8}
  \mathcal{F}(\bm{w}') & =\sum_{i}\mathcal{F}(w'_i)  \notag \\
  &=\sum_{i} \mathcal{F}(w_i) \left[\frac{1} {1+|\mathcal{F}(w_i)|^2\mathcal{F}(n_i)/\mathcal{F}(s_i)}  \right]
\end{align}
The first term $\mathcal{F}(w_i)$ in (\ref{eq:8}) is actually equivalent to the convolution kernel in CNN. Due to the decomposition, the second term $1/ [1+|\mathcal{F}(w_i)|^2 $ $\mathcal{F} (n_i)/\mathcal{F}(s_i) ]$ that represents the influence factor of each kernel can be regarded as working only in a sub frequency band. In this sub-band, we approximate $|\mathcal{F}(w_i)|^2$ in the influence factor to a constant. The strength of the original signal $\mathcal{F}(s_i)$ is also invariable in the task of adapting to different quantization noises, so $|\mathcal{F}(w_i)|^2/\mathcal{F}(s_i)$ can be approximated as a constant $k_i$. For the intensity of the quantization noise $\mathcal{F}(\bm{n})$, it is proportional to the square of Qstep at all frequencies with the default coding setting. Similarly, the decomposed noise $\mathcal{F}(n_i)$ has the same pattern at the selected frequency.

\begin{equation}\label{eq:9}
\frac{|\mathcal{F}(w_i)|^2}{\mathcal{F}(s_i)}\mathcal{F}(n_i)\approx k_i \mathcal{F}(n_i) \propto Q_{step}^2
\end{equation}
We use trainable parameters $\bm{\theta}$ to represent the proportional relationship here. Therefore, (\ref{eq:8}) can be rewritten as follows.
\begin{equation}\label{eq:10}
  \mathcal{F}(\bm{w}')\approx\sum_{i} \mathcal{F}(w_i) \left[\frac{1} {1+\theta_i Q_{step}^2}  \right]
\end{equation}
where the $\theta_i$ indicates a specific parameter in set $\bm{\theta}$.
We assume that $N$ represents the number of feature maps of the CNN, then the number of parameters introduced by this method should be $O(N^2)$, which is the same order as the number of the kernels.
Inspired by DSC \cite{DSC} that uses depthwise convolution instead of standard convolution, we apply the influence factor to the feature maps instead of the convolution kernels. Fig. \ref{fqpm} shows the schematic diagram of our proposed method. Thus the parameter quantity becomes $O(N)$, which is the same order as the number of the biases (the QP map method \cite{song2018practical}).

\subsection{Implementation Detail}
Same as HEVC\cite{quantization}, the relationship between QP and Qstep in VVC can be written as \cite{vtmDescription}:
\begin{equation}\label{eq:id1}
  Q_{step} = 2^{(QP-4)/6}
\end{equation}
so the square of Qstep is:
\begin{equation}\label{eq:id2}
  Q_{step}^2 = 2^{(QP-4)/3}
\end{equation}
Due to the trainable multiplier $\theta_i$ in (\ref{eq:10}), multiplying different $Q_{step}^2$ by the same constant does not affect the performance of the model. A normalization of using $2^{(QP-32)/3}$ to replace $Q_{step}^2$ is adopted, which can avoid the gradient vanishing problem caused by large $Q_{step}^2$.
Besides, the parameter $\theta_i$ should be greater than 0 because both $|\mathcal{F}(w_i)|^2$ and $\mathcal{F}(\bm{s})$ are greater than 0. When $\bm{\theta}$ are $\bm{0}$, the proposed model will turn into the original CNN filter. There are two common methods to solve this: 1. Using the reparametrization of $\bm{\theta} = exp(\bm{\eta})$, where $\bm{\eta}$ are the unconstraint trainable parameters. 2. Directly truncating $\bm{\theta}$. We adopt the second one in this paper.

The order of the filtering process of H.266/VVC with CNN filter is the luma mapping with chroma scaling(LMCS), DB, CNN filter, SAO, and ALF. Both SAO and ALF need to add some bits to indicate the offsets and coefficients. By putting the CNN-based filter before SAO and ALF, a filtered image with higher quality can be sent to SAO and ALF, thereby reducing the number of coded bits. For the DB, which depends on the preset thresholds to perform filtering, putting the CNN filter before it may need to modify its thresholds accordingly. Therefore, instead of using this strategy, we choose to put the CNN filter after it.

\section{Experiment}
In this section, the experimental setting is first introduced.
Then we provide the experiment results on coding efficiency and complexity. Finally, the comparisons with previous work are provided.

\subsection{Experimental Setting}
By integrating our method into different models, the BD-rate \cite{BDrate} reduction of the proposed method could be tested in various situations, such as complexity, activation, serial or parallel structure, etc. Here we chose four different models, including Liu \textit{et al.} \cite{mine}, VRCNN \cite{VRCNN}, DCAD \cite{DCAD}, and Tucodec \cite{Tucodec}, as the backbones for our experiment.
The CNN-filter was integrated into the Versatile Video Coding Test Model (VTM)-6.3 \cite{VTM} and put between DB and SAO.
The DIV2K dataset \cite{DIV2K} was used to train and validate all of the mentioned CNN filters.
We divided 900 pictures in DIV2K into 800 as the training set and 100 as the validation set.
Four QPs including 22, 27, 32, and 37 in common test condition (CTC\cite{CTC}) were used to encode these pictures.
By cutting these pictures into $64 \times 64$ blocks, we obtained 522,877 samples for training and 66,712 samples for validation under each QP. With framework Keras \cite{keras} and optimizer Adam \cite{adam}, about 40,000 iterations were trained for each QP with batch size 128.
In the test phase, the 1-st frames from HEVC test sequences were used for evaluating the performance of the aforementioned filters. It is worth mentioning that these test sequences were not overlapped with the datasets used in the training or validation phase. 

\subsection{Performance Evaluation}\label{perEva}
From the results shown in Table \ref{tab:bdrate},
the "Global" column of using a single model has the lowest BD-rate reduction.
By using multiple models shown in the "Separate" column, the overall BD-rate reduction of the CNN-filter has been improved, but the number of required parameters has increased fourfold.
From the "Proposed" column, it can be seen that our proposed method enables a single model to have excellent performance for all four backbones while only increasing a small number of parameters. This shows the versatility and flexibility of our proposed method. Its BD-rate reduction is almost the same as that of the separate method on the luma component. And it achieves extra 0.2\% BD-rate reduction on the chroma components.
This fully demonstrates that our method effectively improves the generalization ability of the model, because we only use the luma component for training. The performance of the chroma components is completely dependent on the generalization ability of the model.
As shown in Fig. \ref{fig:psnr}, the PSNR gains of the methods relative to the VTM baseline are also plotted. The dotted line represents the separated method, and the QP in its legend represents the dataset used for training. We can find that the lines peak in the QP of the corresponding training dataset but perform poorly on the other QPs. Especially for lower QP, it may even lead to a negative impact on performance. The blue solid line represents the global method, which can obtain an ordinary BD-rate gain at higher QP, but similarly, it hurts the reconstructed image at lower QP.
On the contrary, the proposed method has a significant filtering performance in a wide range of QPs and almost reaches the optimal performance of using multiple separate models. This further demonstrates the effectiveness and versatility of our model.

\begin{table}[!tbp]
  \footnotesize
  \centering
  \caption{\small Comparison of Relative Decoding Complexity}
    \setlength{\tabcolsep}{1mm}{
    \begin{tabular}{c|r|r|r|r|r|r}
    \hline
    \multirow{2}[4]{*}{Class} & \multicolumn{2}{c|}{Liu \textit{et al.} \cite{mine}} & \multicolumn{2}{c|}{DCAD \cite{DCAD}} & \multicolumn{2}{c}{Tucodec \cite{Tucodec}} \bigstrut\\
\cline{2-7}      & \multicolumn{1}{c|}{Global} & \multicolumn{1}{c|}{Proposed} & \multicolumn{1}{c|}{Global} & \multicolumn{1}{c|}{Proposed} & \multicolumn{1}{c|}{Global} & \multicolumn{1}{c}{Proposed} \bigstrut\\
    \hline
    A & 345.3\% & 353.4\% & 602.1\% & 612.5\% & 683.2\% & 694.6\% \bigstrut\\
    \hline
    B & 453.4\% & 461.0\% & 693.1\% & 704.9\% & 794.6\% & 814.5\% \bigstrut\\
    \hline
    C & 432.2\% & 442.7\% & 775.4\% & 770.1\% & 817.3\% & 834.0\% \bigstrut\\
    \hline
    D & 585.3\% & 627.4\% & 1443.2\% & 1461.3\% & 1463.7\% & 1487.9\% \bigstrut\\
    \hline
    E & 555.3\% & 563.8\% & 1047.2\% & 1059.8\% & 1148.0\% & 1177.0\% \bigstrut\\
    \hline
    Average &    \textbf{474.3\%} & \textbf{489.6\%} & \textbf{912.2\%} & \textbf{921.7\%} & \textbf{981.4\%} & \textbf{1001.6\%} \bigstrut\\
    \hline
    \end{tabular}%
    }
  \label{tab:complexity}%
\end{table}%

\subsection{Complexity Evaluation}
In Table \ref{tab:bdrate}, the parameter comparisons of different methods are shown in the "Param." column. In addition, Table \ref{tab:complexity} shows comparisons of relative decoding complexity than VTM anchor, which shares the same test setting with Section \ref{perEva}.
Although the separate method uses more models than the global method, for the filtering of determined QP, they both use a single model with the same structure, so their complexity should be the same.
From Table \ref{tab:complexity}, the decoding complexity of our proposed method only increases by about 2\% compared with the global method. Therefore, the impact of our proposed method on complexity is minimal. This lays a good foundation for the practical application of our method.

\begin{figure}
  \centering
  \includegraphics[width=8.5cm]{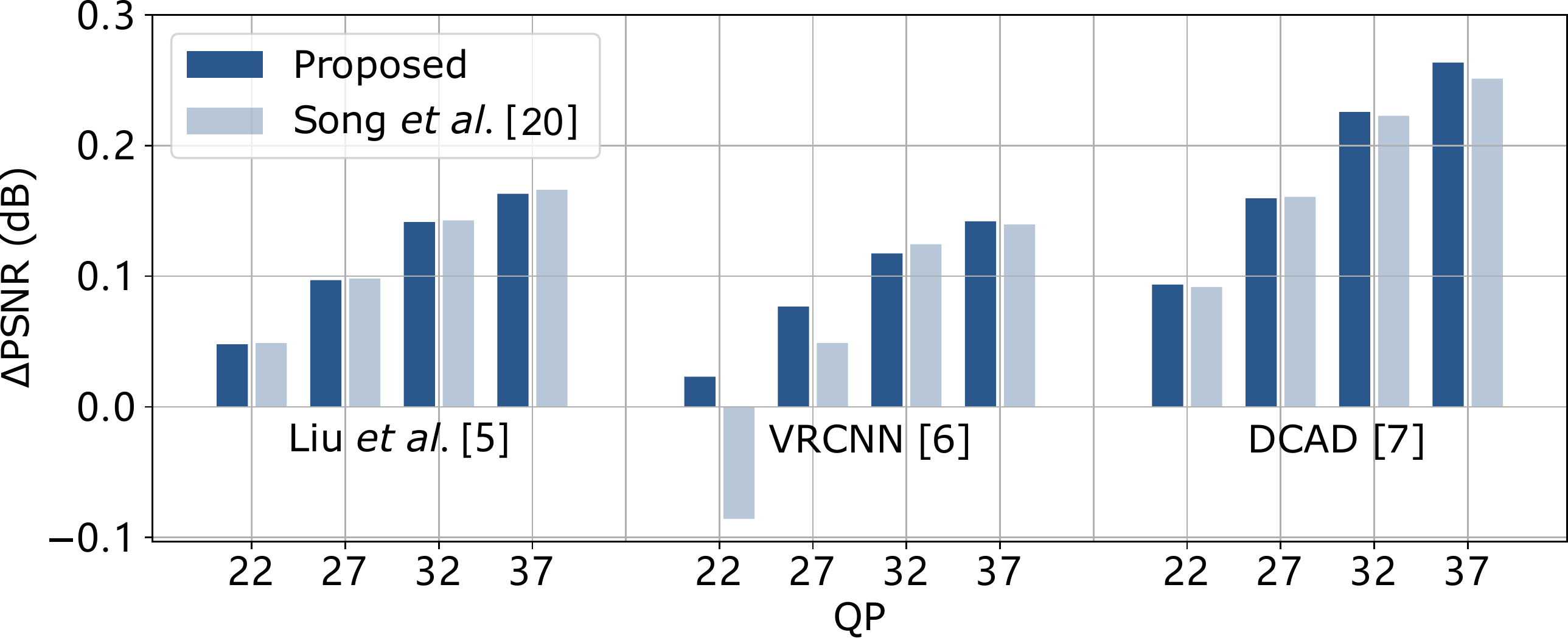}
  \caption{\small Comparison of the overall relative PSNR gain of Song \textit{et al.} \cite{song2018practical} and the proposed method. }\label{compare_ref}
\end{figure}

\subsection{Comparison with Previous Work}
The performance comparison of the proposed method and the QP map method (Song \textit{et al.}\cite{song2018practical}) are shown in Fig. \ref{compare_ref}. These two methods achieve similar relative PSNR gain with the backbone of Liu \textit{et al.} \cite{mine}, DCAD \cite{DCAD}. For VRCNN\cite{VRCNN}, Song \textit{et al.}\cite{song2018practical} has a negative impact on lower QPs, whereas our proposed method still achieves a minor gain when $QP=22$. Our method performs more robust, probably because it provides the quantization information for each convolution layer but Song \textit{et al.}\cite{song2018practical} only does it for the input one.
This comparison demonstrates the robustness and versatility of our method.

\section{Conclusion}
In this paper, we present a novel method to improve the adaptability of CNN-filters to different QPs.
By adding influence factors related to Qstep to the CNN-filter, CNN can suppress the quantization noise as the noise changes.
The proposed method achieves excellent performance on previous CNN-filters and yields similar BD-rate reduction to using multiple models. Besides, the complexity evaluations of different trained models prove that it only brings a slight increase in complexity and has a promising future for practical applications. Finally, the comparison with previous work shows that our proposed method is more robust and stable. We believe that in the future, more efficient methods will emerge based on further design and modeling.

\section{Acknowledgment}

This work was supported in part by the National Natural Science Foundation of China under Grant 61674041, in part by Alibaba Group through Alibaba Innovative Research (AIR) Program, in part by the STCSM under Grant 16XD1400300, in part by the pioneering project of academy for engineering and technology and Fudan-CIOMP joint fund, in part by the National Natural Science Foundation of China under Grant 61525401, in part by the Program of Shanghai Academic/Technology Research Leader under Grant 16XD1400300, in part by the Innovation Program of Shanghai Municipal Education Commission, in part by JST, PRESTO Grant Number JPMJPR19M5, Japan.


\bibliographystyle{IEEEbib}
\bibliography{reference}

\end{document}